\documentclass[%
reprint,
superscriptaddress,
amsmath,amssymb,
aps,
prb,
floatfix
]{revtex4-1}

\usepackage{multirow}
\usepackage{graphicx}
\usepackage{dcolumn}
\usepackage{bm}
\usepackage[version=3]{mhchem}
\usepackage{nicefrac}

\usepackage{color}
\usepackage{dcolumn}
\usepackage{amssymb}
\usepackage{url}
\usepackage{hyperref}
\usepackage{xfrac}
\usepackage{tabularx}
\usepackage{xspace}
\usepackage{amsmath}
\usepackage[normalem]{ulem}
\usepackage{mathtools}
\usepackage{caption}
\usepackage{subcaption}

\def\be{\begin{equation}}
\def\ee{\end{equation}}
\def\bea{\begin{eqnarray}}
\def\eea{\end{eqnarray}}
\def\ba{\begin{array}}
\def\ea{\end{array}}

\def\MODELNAME{Model-X\xspace}

\begin{document}
\title{Exact quantum ground state of a two-dimensional quasicrystalline antiferromagnet}
\author{Pratyay Ghosh}
\email{pratyay.ghosh@uni-wuerzburg.de}
\affiliation{Institut f\"ur Theoretische Physik und Astrophysik and W\"urzburg-Dresden Cluster of Excellence ct.qmat, Universit\"at W\"urzburg,
Am Hubland Campus S\"ud, W\"urzburg 97074, Germany}


\begin{abstract}
We present the exact dimer ground state of a quantum antiferromagnet defined on a quasicrystal constructed from the Bronze-mean hexagonal quasicrystal. A coupling isotropy on the first and second-neighbor bonds is sufficient to stabilize a product state of singlets on the third-neighbor bonds. We also provide a systematic approach for constructing additional crystals, quasicrystals, and amorphous structures that can sustain an exact dimer ground state.
\end{abstract}

\maketitle

\section{Introduction} The quantum antiferromagnets on two-dimensional frustrated lattices have been a focus of condensed matter research for decades~\cite{frustrationbook,diepbook}. Owing to the non-commutativity of quantum spin operators and the frustrated magnetic interactions, it is often impossible to find an analytic solution to these systems' ground states. So, while dealing with these systems, one usually analyzes the possible ground states allowed by the point- and space-group symmetries of the underlying lattices to gain qualitative understandings or to perform advanced numerical investigations~\cite{Wen1989,Wen2002,Messio2011}; in all these analyses, the lattice translations play a pivotal role. With the appearance and increasing number of quasicrystalline materials~\cite{Shechtman1984,Watanuki2012,Ishikawa2016,Sato2019,Miyazaki2020}, however, it became apparent that the spin systems may not always be periodic~\cite{Janot1992}, leading the frustrated quasicrystals to frequently realize exotic ground states, such as spin-glass~\cite{Fukamichi1987,Berger1990,Islam1998,Ishikawa2016}, which are hard to decipher due to the unavailability of lattice periodicity.

From our experience with crystalline quantum magnets, we may infer that the leading-edge knowledge for understanding complex spin systems is often provided by exactly solvable models. Shastry and Sutherland proposed the first such exactly solvable model in 2D~\cite{Shastry1981}. The Shastry-Sutherland model (SSM), defined on a lattice of uniform tiling, was found to exhibit an exact dimer singlet ground state. Though such exact dimer ground states are limited only to two 2D lattices with uniform tiling~\cite{Ghosh2022} (maple-leaf model (MLM) is the other one), they have widely served as crystallization seeds for several theoretical and experimental advancements. Despite the importance, such exact solutions in quasicrystals are very limited~\cite{Korepin1987,Jeon2022}, although there have been an extensive number of works on quasicrystalline spin systems ~\cite{Hauser1986,Oxborrow1993,Wessel2003,Wessel2005,Jagannathan2005,Jagannathan2007}, due to the complexity of the aperiodicity.

In this letter, we propose a model defined on a two-dimensional quasicrystal that admits an exact dimer ground state. We, first, construct the quasicrystal from the bronze-mean hexagonal quasicrystal (BMHQ)~\cite{Dotera2017}, via site and bond depletion. Next, we include a subset of third-neighbor couplings. By demonstrating that this model has an exact dimer ground state, a product state of dimer singlets, we, then, assess its stability. In the end, we introduce a generic method, that equally applies to crystals, quasicrystals, and amorphous systems, which creates further models bearing exact dimer ground states.

\begin{figure*}
\includegraphics[width=0.85\textwidth]{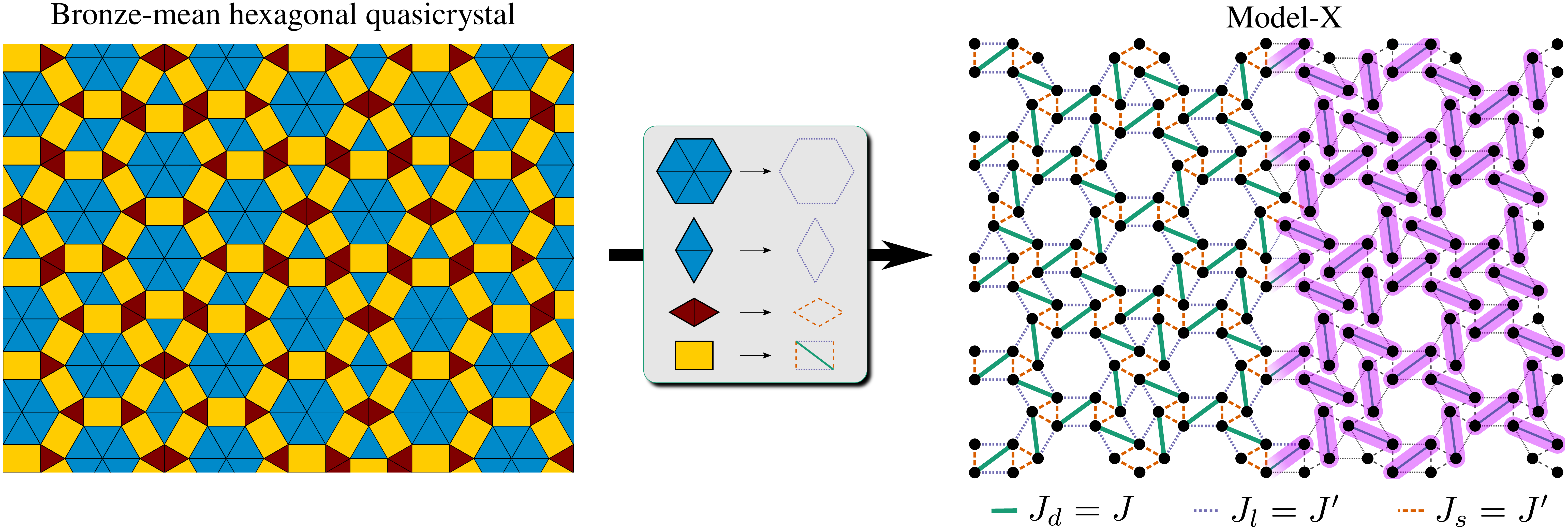}
\caption{\textbf{The construction of \MODELNAME from Bronze-mean hexagonal quasicrystal:} First, we deplete all vertices that are shared by six big triangles to generate the hexagonal tiles. Next, we deplete the connections shared by two triangles, constructing the two rhombus tiles. On the resulting tiling, we place a spin on each vertex, and the connection between the vertices, which now connects a pair of spins, acts as the bonds. The bonds are of two different lengths, namely $s$ and $l$, shown by dashed orange and dotted blue lines. We assume an equal Heisenberg coupling of strength $J'$ on both these bonds. We further add a subset of the third neighbor interactions, shown by thick green lines, with an exchange interaction of strength of $J$. We call the resulting spin-model \MODELNAME. The singlets (light purple ellipses) reside on $J$ bonds. As the $J'$ couplings do not contribute to the energy, this product state of singlets is an eigenstate of \eqref{eq-ham1}.} \label{fig-1}
\end{figure*}

\section{Model}
To construct our model, which hosts an exact dimer state, we start with the BMHQ~\cite{Dotera2017} (see Fig.\ref{fig-1}), which is closely related to the celebrated Penrose tiling~\cite{Penrose1974} and the Ammann-Beenker tiling~\cite{Grunbaum1977}, in a sense, as their inflation ratios are all metallic ratios -- for the BMHQ it is the bronze ratio, and for the other two, it is the golden, and silver ratio, respectively. The BMHQ consists of three different tiles: small equilateral triangles (edge length $s$), big equilateral triangles (edge length $l$), and $s\times l$ rectangles. A six-fold-symmetric irregular dodecagon constructed of six copies of each tile is the elementary motif of BMHQ. The discussion in Ref.~\cite{Ghosh2022} makes it apparent that the BMHQ is incapable of supporting an exact dimer eigenstate if we place a spin on every vertex (site) and couple them pair-wise via Heisenberg exchange interactions (bonds). To host an exact dimer state, the system must satisfy a necessary, but not sufficient, condition of having an odd coordination number, which is not the case for BMHQ. Therefore, we perform the following. We deplete all sites shared by six adjacent large triangles, and all bonds shared by two triangles. As a result, we get a different quasicrystal (depicted in Fig.\ref{fig-1}), consisting of the six tiles: hexagons (edge length $l$), small rhombi (edge length $s$), large rhombi (edge length $l$), small triangles (edge length $s$), large triangles (edge length $s$), and $s\times l$ rectangles. Lastly, we add half of the third-neighbor couplings, i.e. one diagonal of each rectangle, such that every site is only part of one diagonal interaction. Note that there are two sets of diagonal bonds one can choose from; both are adequate for our purpose.

We now define a Heisenberg spin Hamiltonian on this quasicrystal as
\be\label{eq-ham}
\mathcal{H}=J_s\sum_{\mathclap{\langle kl\rangle}} \vec{S}_k\cdot\vec{S}_l+J_l\sum_{\mathclap{\langle\langle km\rangle\rangle}}\vec{S}_k\cdot\vec{S}_m+J_d\sum_{\mathclap{( lm)}}\vec{S}_l\cdot\vec{S}_m.
\ee
Here, $\vec{S}_i$ denotes the $\mathfrak{su}(2)$ spin-$S$ operator at site $i$, $\langle kl\rangle$, $\langle\langle km\rangle\rangle$, and $(lm)$ runs over orange dashed, blue dotted, and thick green bonds, respectively (Fig.~\ref{fig-1}). Here, we only discuss the special case $J_d=J$ and $J_s=J_l=J'$ of~\eqref{eq-ham},
\be\label{eq-ham1}
\mathcal{H}_X=J'\sum_{\mathclap{\langle kl\rangle}} \vec{S}_k\cdot\vec{S}_l+J'\sum_{\mathclap{\langle\langle km\rangle\rangle}}\vec{S}_k\cdot\vec{S}_m+J\sum_{\mathclap{( lm)}}\vec{S}_l\cdot\vec{S}_m,
\ee
(henceforth dubbed as Model-X), which can admit an exact dimer ground state.

\section{Quantum Ground State}
The exact dimer ground state of \MODELNAME can be obtained by following the same procedure as in Refs.~\cite{Shastry1981,Ghosh2022}. First, we recast~\eqref{eq-ham1} into a sum over the interacting spins on the right-angled triangles as
\be\label{eq-hamil-tri}
\mathcal{H}_X=\sum\left(\begin{gathered}\includegraphics[scale=0.3]{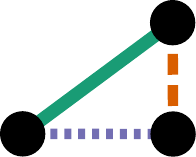}\end{gathered}+\begin{gathered}\includegraphics[scale=0.3,angle=60,origin=r]{./t1.pdf}\end{gathered}+\begin{gathered}\includegraphics[scale=0.3,angle=120,origin=r]{./t1.pdf}\end{gathered}+\begin{gathered}\includegraphics[scale=0.3,angle=180,origin=r]{./t1.pdf}\end{gathered}+\begin{gathered}\includegraphics[scale=0.3,angle=240,origin=r]{./t1.pdf}\end{gathered}+\begin{gathered}\includegraphics[scale=0.3,angle=300,origin=r]{./t1.pdf}\end{gathered}\right).
\ee
For this, we have distributed the interactions $J\vec{S}_l\cdot\vec{S}_m$ on the thick green bonds in Fig.~\ref{fig-1} equally between the two triangles that share this bond. Thus, these triangles now have three different colored bond interactions, the Hamiltonian for each, in general, reads as
\be\label{eq-ham2}
h_{\triangle}=J' \vec{S}_k\cdot\vec{S}_l+J'\vec{S}_k\cdot\vec{S}_m+\frac{J}{2}\vec{S}_l\cdot\vec{S}_m.
\ee
The triangular decomposition of~\eqref{eq-ham1}, in a sense, sculpts the model into a frustration-free form~\cite{Majumdar1969,Shastry1981,AKLT,Klein_1982}, as the ground state minimizes the energy of each $h_\triangle$. Note that if there are $N$ spins in our model, there will be $N$ such right-angled triangles, and $N/2$ thick green bonds.

When $J/2>J'$, the ground state of $h_\triangle$ is a spin-singlet forming on the green bond, which we denote by $\big| [lm] \big\rangle$. Now, the parity of the singlet causes the first two terms in~\eqref{eq-ham2} to cancel each other's contribution to the energy of a triangle, i.e. $$(J' \vec{S}_k\cdot\vec{S}_l+J'\vec{S}_k\cdot\vec{S}_m)\big| [lm] \big\rangle=0.$$ Therefore, we can construct a product state of $\big| [lm] \big\rangle$ which covers the entire system as
\be
|\psi\rangle=\otimes\underset{\mathclap{( lm )}}{\prod} \big| [lm] \big\rangle, \label{eigen}
\ee
(see Fig.\ref{fig-1} where we overlay $|\psi\rangle$ on \MODELNAME). Again, due to the parity of the $\big| [lm] \big\rangle$s, the first two terms of~\eqref{eq-ham1} do not contribute to the energy of the full system or renormalize \eqref{eigen}, thus, making it an exact eigenstate of \MODELNAME with an energy density, which is independent of $J'$, given by
\be
E/N= -\frac{S(S+1)}{2}J.\label{energ}
\ee

If the ground state energy of~\eqref{eq-ham2} is $e_\triangle$, then~\eqref{energ} sets an upper bound for the ground state energy of the entire system, i.e., $E_g/N\geq e_\triangle$, the equality of which holds when $J$ is greater than a lower bound $J_{b}$ and $|\psi\rangle$ is the ground state. For spin-$1/2$, $J_b$ can be easily computed to be $2J'$, and can likewise be obtained for other spins~\footnote{Here, we only covered Heisenberg spin interactions. In general, an analysis along these lines can also be achieved for XXZ-type interactions and a Zeeman term. For further details, see Refs.~\cite{Shastry1981, Ghosh2022}}. Note that the variational principle-based analysis that we have performed thus far, does not prevent $|\psi\rangle$ from being the ground state for $J<J_b$, only it cannot be shown analytically. The critical $J_c$, such that for $J_b>J\ge J_c$ the exact dimer state is the ground state of the system, can only be obtained numerically~\cite{Corboz2013,Lee2019,Farnell2011,Ghosh2022}, which we attempt next.
\begin{figure}
\includegraphics[width=0.6\columnwidth]{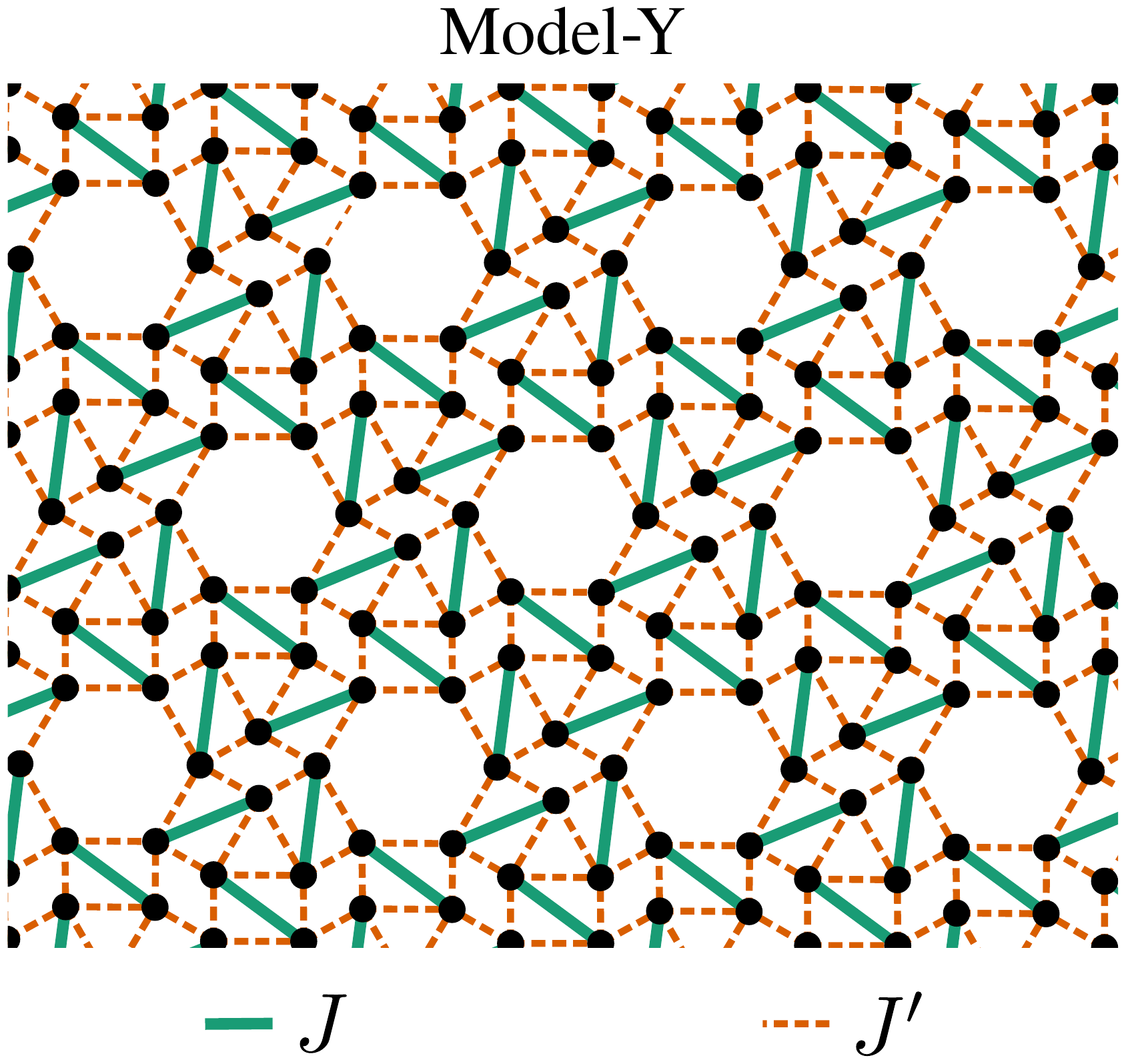}
\caption{The Model-Y, defined on a lattice, is made out of the same tiles as the \MODELNAME. In Model-Y, the dashed orange bonds have an exchange interaction of strength $J'$, and the green bonds bare an exchange interaction of strength $J$. Similar to \MODELNAME, Model-Y can also admit an exact singlet eigenstate with singlets on the $J$ bonds, which becomes a ground state of the system for $J\gtrsim1.45J$.} \label{fig-2}
\end{figure}

\section{Stability of the Exact Dimer Phase}
Due to the aperiodicity and frustration, \MODELNAME is impregnable by most numerical techniques in their status quo. Therefore, we resort to the robust density matrix renormalization group (DMRG) approach to obtain an estimate of $J_c/J'$~\cite{White1992}. We perform our DMRG calculations on a 144-site spin-$1/2$ cluster (refer to Appendix for more details) using the ITensor library~\cite{Fishman2020}. The critical $J_c/J'$ found from our calculations is shown in Tab.~\ref{tab-1}. The phase transitions out of the dimer phase in both SSM and MLM have been found to be first-order in nature~\cite{Corboz2013,Lee2019,Farnell2011}. One may wonder if the aperiodicity for \MODELNAME can change this nature or if the system can form domains of exact dimer and non-dimer states. This seems implausible because when a local dimer state switches to a non-dimer state, the system's structure causes the dimers nearby to experience local non-uniform effective fields that force them to also become non-dimer states. As a result, a chain reaction spreads across the whole system, thereby, making the product singlet state kaput as a whole. Thus, we do not expect a second-order phase transition out of the exact dimer phase in \MODELNAME, which is confirmed by our DMRG calculations. 

To gain further insights, we also compare our current dimer phase's stability with the same in SSM and MLM in Tab.~\ref{tab-1}, with intra- and inter-dimer coupling being $J$ and $J'$, respectively. We find \MODELNAME to have a comparatively less stable exact dimer phase. To make a further comparison, we introduce a lattice made out of the same tiles (except the triangles of edge $s$) as \MODELNAME, shown in Fig.~\ref{fig-2} (henceforth called Model-Y). Model-Y also admits a product singlet ground state on the $J$ dimers. Here, we again take a 144-site spin cluster and perform DMRG calculations to study the stability of its exact dimer phase. The result is shown in Tab.~\ref{tab-1}. A comparison of the number of triangular and quadrangular tiles makes it apparant that magnetic frustrations in \MODELNAME and Model-Y are more than SSM but less than MLM. As the frustration impacts the exact dimer phase's stability~\cite{Ghosh2022}, one can anticipate that $J_c$ for \MODELNAME and Model-Y both will fall somewhere between the same for SSM and MLM. Model-Y exhibits this, but \MODELNAME does not. The most likely explanation could be that the DMRG results on a finite section of quasicrystal, where the local structure might severely affect the exact dimer state, are very different from the results at the thermodynamic limit, where such local effects are averaged out. However, we can not exclude the possibility that the system's aperiodicity might play a role in destabilizing the exact dimer state, even at the thermodynamic limit. A thorough stability analsysis of the exact dimer phase is beyond the scope of this letter; via our DMRG calculations, we only demonstrate that the exact dimer state on this quasicrystal can be stable even for $J<2J'$.

Though we do not investigate the models in detail beyond their exact dimer ground state, we can anticipate the possibility that \MODELNAME and Model-Y can have other novel phases in the high frustration regime $J<J_c$ (similar to SSM and MLM~\cite{Koga2000,Corboz2013,deconfined,Shi2022,Yang2022,Farnell2011}). This makes both our models worthy of further investigations, e.g. the nature of these additional phases and the phase transitions which might feature the exotic deconfined criticality~\cite{deconfined,Yang2022}.
\begin{table}[]
\begin{tabular}{|p{0.2\linewidth}|p{0.2\linewidth}|p{0.2\linewidth}|}
\hline
& $J_c/J'$& Refs. \\ \hline
MLM & $1.35$&Ref.~\cite{Ghosh2022} \\ \hline
Model-Y & $1.45(1)$& This work \\ \hline
SSM & $1.48$&Ref.~\cite{Corboz2013,Lee2019} \\ \hline
Model-X & $1.49(1)$& This work \\ \hline
\end{tabular}
\caption{The comparison of the stability of the exact dimer ground state in different models. For $J>J_c$, the exact dimer states of the corresponding models become their ground state. For MLM, Ref.~\cite{Farnell2011} finds a $J_c/J'\approx1.45$.} \label{tab-1}
\end{table}

\begin{figure*}
\includegraphics[width=0.9\textwidth]{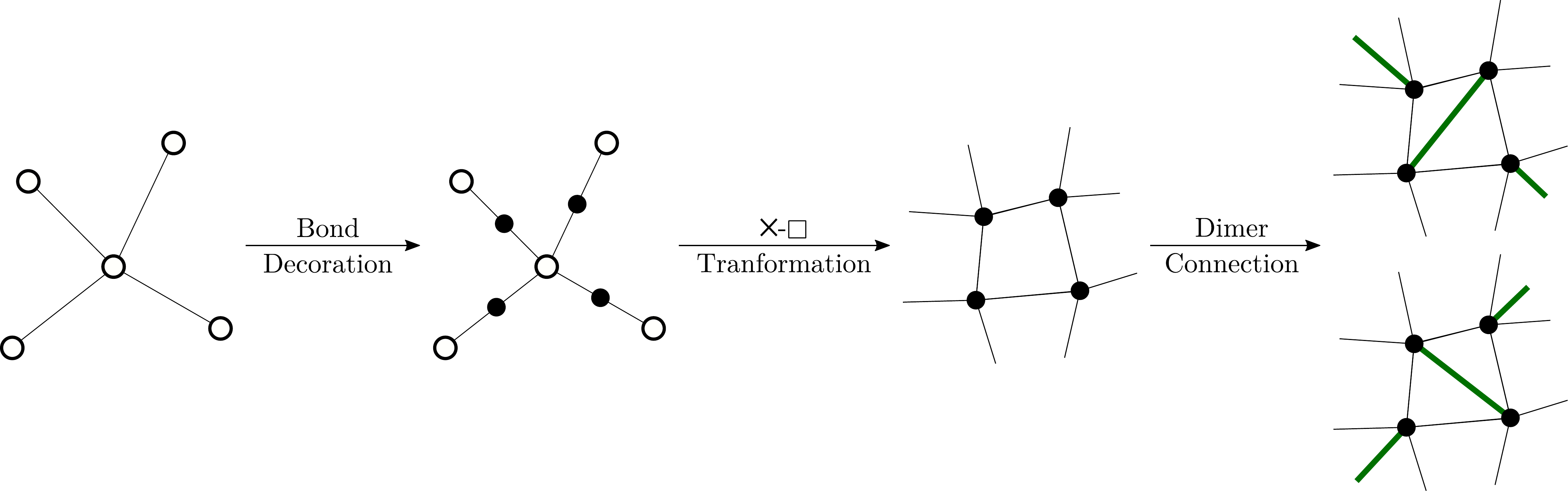}
\caption{A general scheme for obtaining a model with an exact dimer ground state from a graph where each vertex has four connections. First, we decorate the connections with new vertices (marked by filled circles), which is followed by a generalized star-triangle-type transformation to decimate all the old vertices. From there, one can generate two possible graphs by connecting a subset of the diagonals of the newly generated quadrangles (keeping in mind that no site can be part of two diagonals), both of which can host an exact dimer state.} \label{fig-3}
\end{figure*}
\section{Further Possibilities}
We have introduced \MODELNAME, defined on one quasicrystal, which admits an exact dimer ground state. However, the question remains: Are there other quasicrystals with exact dimer ground states? The answer is yes, and we are now going to outline how to construct a specific class of such systems, i.e. with a coordination number five. It is also possible to construct such systems with any odd (homogeneous and mixed) coordination numbers. Since the prescription we present here applies to graphs, it is suitable for all types of systems -- crystalline, quasicrystalline, and amorphous. We begin with a planar graph where each vertex is connected to four other vertices (see Fig.~\ref{fig-3}). In the next step, the connections are all decorated by adding a new vertex (the filled circles in Fig.~\ref{fig-3}). These new vertices will be our actual sites carrying the spins. After that, one applies a generalized version of star-triangle-type transformation (we deem it as $\times$-$\square$ transformation) to decimate the original vertices and form connections between the new vertices. In our spin model these new connections act as inter-dimer bonds. At this point, we have produced a graph where each vertex is part of two generalized quadrangles. In our final step, we connect one diagonal of each quadrangle, which serves as our dimer bond, while ensuring that any two dimers do not share a site. Thus, we construct a system that can host an exact dimer ground state, which is the product state of singlets on the dimer bonds, and the proof is similar to SSM~\cite{Shastry1981}, MLM~\cite{Ghosh2022}, and \MODELNAME. In the simplest case, one assigns equal strength to all inter-dimer couplings. All the intra-dimer couplings also have equal strength but are different from the inter-dimer couplings. However, more complex models can also be defined on such a graph that allow exact dimer states. Note that the last step can result in two independent graphs (see the last panel of Fig.~\ref{fig-3}), both ideal for our purpose.

Taking the square and the kagome lattice as examples of four coordinated lattices, and then following our procedure one obtains the SSM and the MLM, respectively. To construct such quasicrystals and amorphous systems, however, the initial difficulty is to acquire a system with coordination number 4 on which our prescription can be implemented. This is simple for amorphous systems. One can draw random straight lines on a plane and, in general, this would result in a system that has a coordination number of 4 when one considers the intersections as vertices and the line segments between them as edges. For quasicrystals, one can start with an existing quasicrystal made up of quadrangles, e.g. the Penrose rhomb tiling~\cite{Penrose1974,Grunbaum1977}, place a vertex in the middle of each tile (the dual lattice~\cite{Dotera1990}), and then connect all pairs of vertices if their corresponding tiles share an edge, and thus, one can obtain a 4-coordinated quasicrystal (refer to Appendix for more details).

\section{Conclusion and Outlook}
We have introduced \MODELNAME, which is defined on a quasicrystal made out of hexagons, rhombi, triangles, and rectangles, and studied its exact dimer ground state. To the best of our knowledge, no such model on quasicrystals has been reported ever to exhibit such a property. We also create a crystal using the same tiles and do similar studies on that as well. Finally, we lay out a general scheme for constructing crystals, quasicrystals, and amorphous systems, that can admit an exact dimer ground state.

The \MODELNAME opens up several questions which require further investigations. First, one needs to understand how the aperiodicity influences the stability of the dimer state \eqref{eigen}. Second would be the study of the nature of the other phases, and the possible phase transitions in \MODELNAME, also with the bond anisotropy $J_s\neq J_l$. The third is the investigation of \MODELNAME in a finite magnetic field. The lattice versions of exact dimer models, e.g. the SSM and the MLM, show a series of spin-density wave and multi-triplet bound-state crystal-based magnetization plateaux~\cite{Momoi2000,Fukumoto2000,Miyahara2003,Dorier2008,Corboz2014,Ghosh2023}, behind all of which the lattice periodicity plays a pivotal role. One can still speculate the formation of two and three-triplet bound states in \MODELNAME. However, how the aperiodicity of the model would affect the magnetization process in this system is an extremely tempting question. Lastly, a material realization of \MODELNAME will be highly sought out for, in general, similar to the experimental realizations of SSM, which have played a central role in numerous theoretical and experimental developments. Additionally, our scheme for creating systems with exact dimer ground states will significantly advance the study of amorphous spin systems, a subject in which the exact solution has just lately begun to emerge~\cite{Cassella2022}.

\section{Acknowledgments} The author acknowledges Tobias M\"uller and Ronny Thomale for useful discussions. The work is supported by the Deutsche Forschungsgemeinschaft (DFG, German Research Foundation) through Project-ID 258499086-SFB 1170 and the Würzburg-Dresden Cluster of Excellence on Complexity and Topology in Quantum Matter – ct.qmat Project-ID 390858490-EXC 2147.

\appendix
\subsection*{Appendix A: Details of DMRG calculations}\label{app:DMRG}
\begin{figure}[]
\centering
\begin{subfigure}[b]{0.45\textwidth}
\caption{}\label{fig-S1}
\centering
\includegraphics[width=\textwidth]{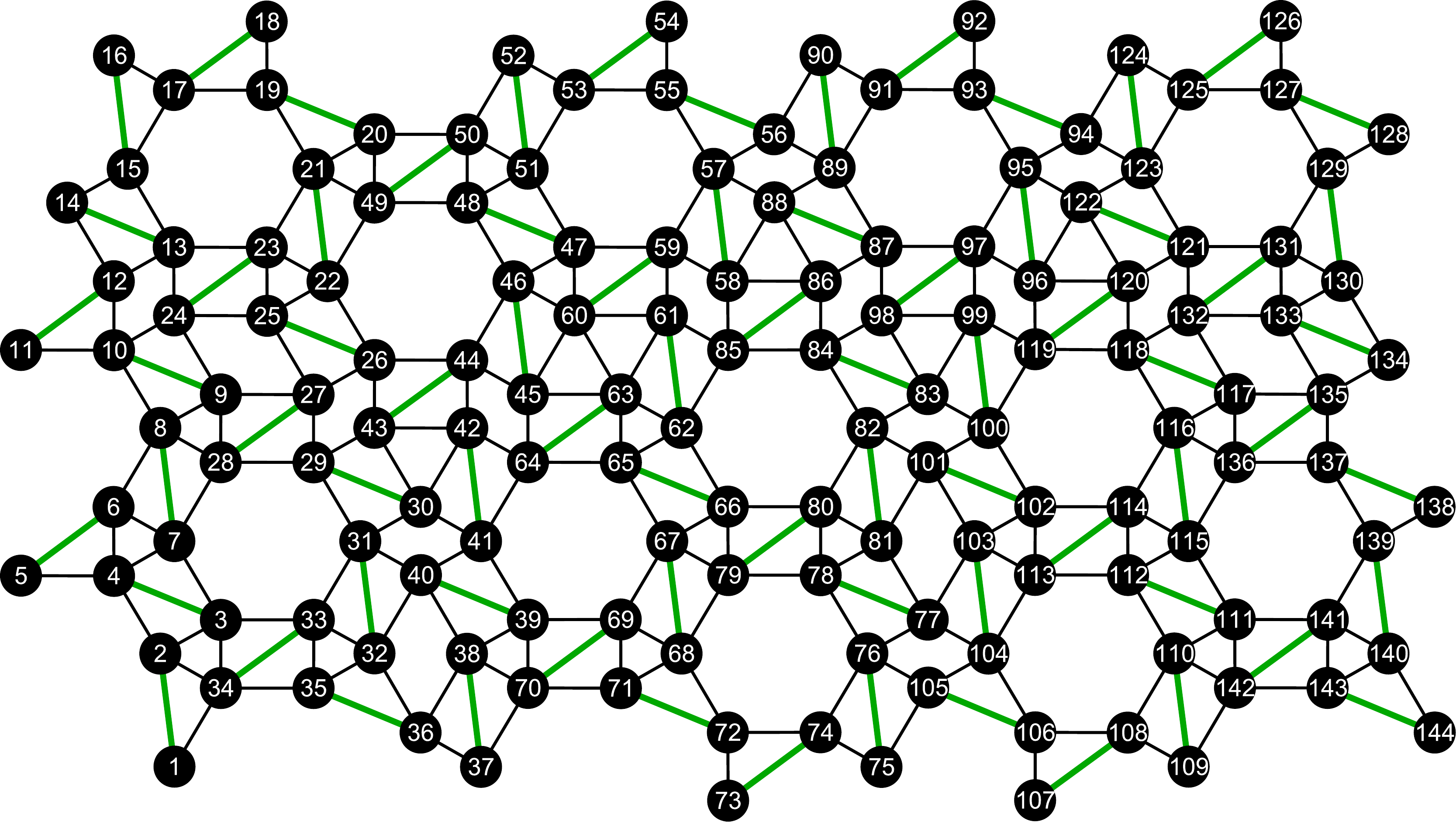}
\end{subfigure}
\vfill
\begin{subfigure}[b]{0.475\textwidth}
\caption{}\label{fig-S2}
\centering
\includegraphics[width=\textwidth]{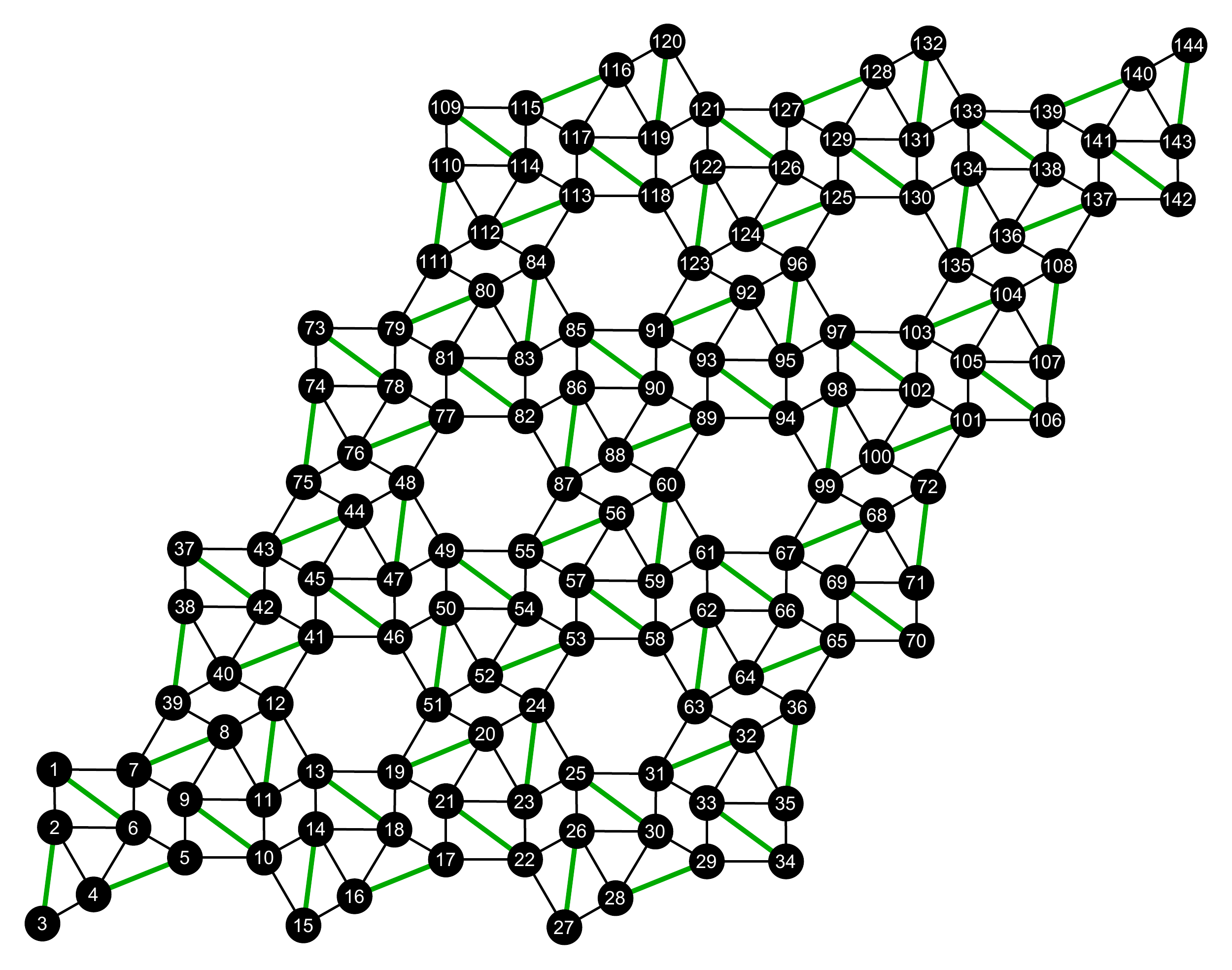}
\end{subfigure}
\caption{(a) The 144-site Model-X cluster used to perform the DMRG calculations mentioned in the main text. (b) The 144-site Model-Y cluster used to perform the DMRG calculations mentioned in the main text.}
\label{fig:child_23 (RIght)}
\end{figure}

\begin{figure*}
\includegraphics[width=0.9\textwidth]{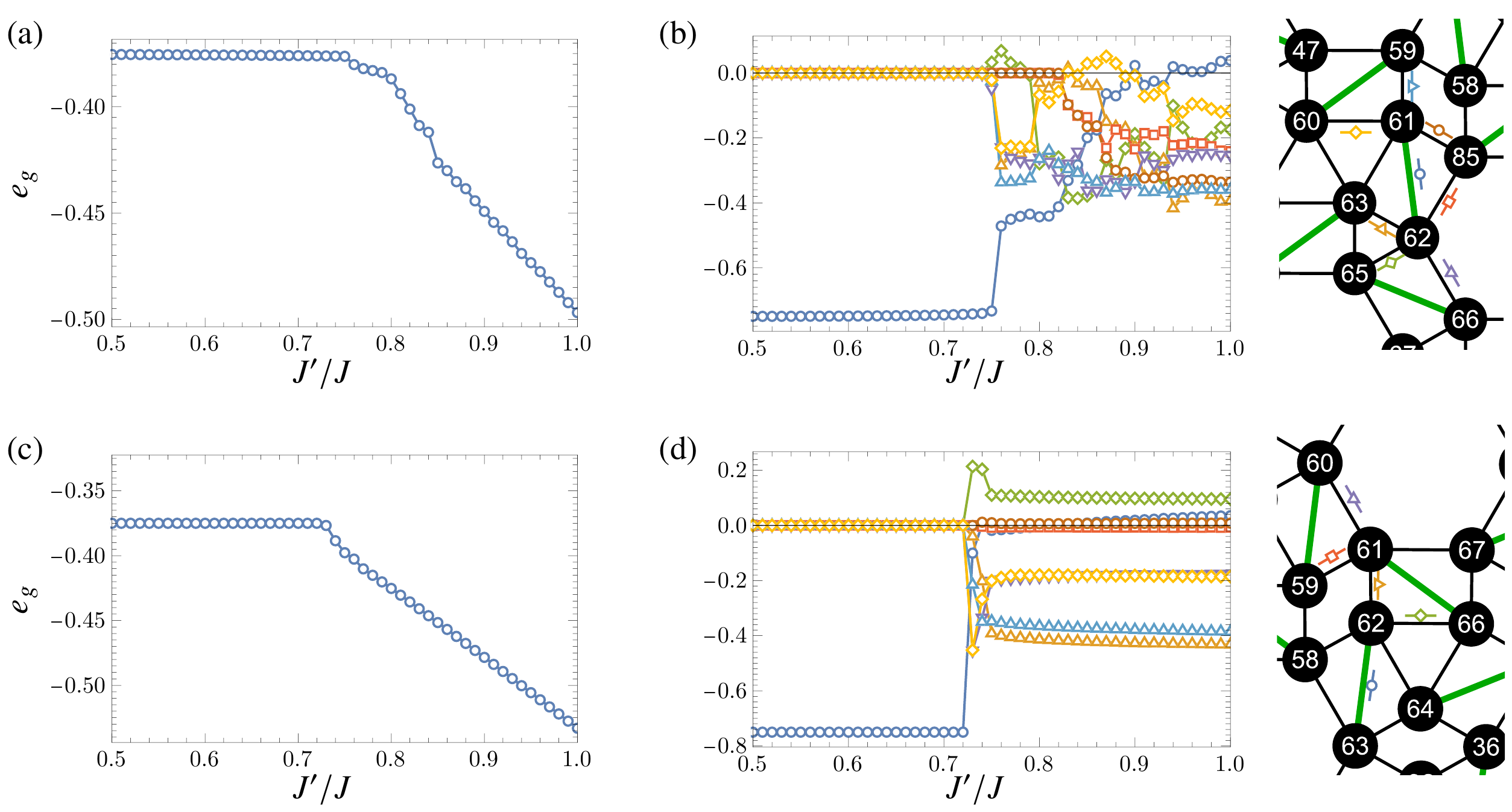}
\caption{The DMRG results for Model-X: (a) the ground state energy per site and (b) the spin-spin correlations on selected bonds. (c) and (d) are respectively the same for Model-Y. The spin systems used for our calculations are depicted in Figs.~\ref{fig-S1} and ~\ref{fig-S1}. In both calculations we have used a open boundary condition for a fair comparison.} \label{fig-S3}
\end{figure*}

Our DMRG calculations which are performed using iTENSOR library~\cite{Fishman2020}. For both model-X and Y we choose a 144 site cluster (see model-X in Fig.~\ref{fig-S1} and model-Y in Fig.~\ref{fig-S2}) with open boundary and perform 24 sweeps with a maximum bond dimension of 1024. The results, i.e. the ground state energy and the spin-spin correlations on a few selected bonds, are shown in Fig~\ref{fig-S3}, which shows a clear first order transition out of the exact dimer phase. {It should be noted that the exact dimer state is only short-range entangled, and while doing DMRG, we have carefully indexed our sites such that no dimer becomes long-ranged due to the effective mapping to a 1D problem. However, when the system exits the exact dimer phase, the long-range entanglements would start to develop. As DMRG suffers from a bias towards less entangled states, it can slightly overestimated the stability of the exact dimer state.} 

\subsection*{Appendix B: Example of a amorphous system with exact ground state}\label{app:am}
Fig.~\ref{fig-S4} shows a amorphous system with a exact dimer ground state which a product of singlets on the thick green bonds. This is created by using the procedure explained in the main text.

\subsection*{Appendix C: Example of a four coordinated quasicrystal}\label{app:qc}
\begin{figure*}[]
\centering
\begin{subfigure}[b]{0.465\textwidth}
\caption{}\label{fig-S4}
\centering
\includegraphics[width=\textwidth]{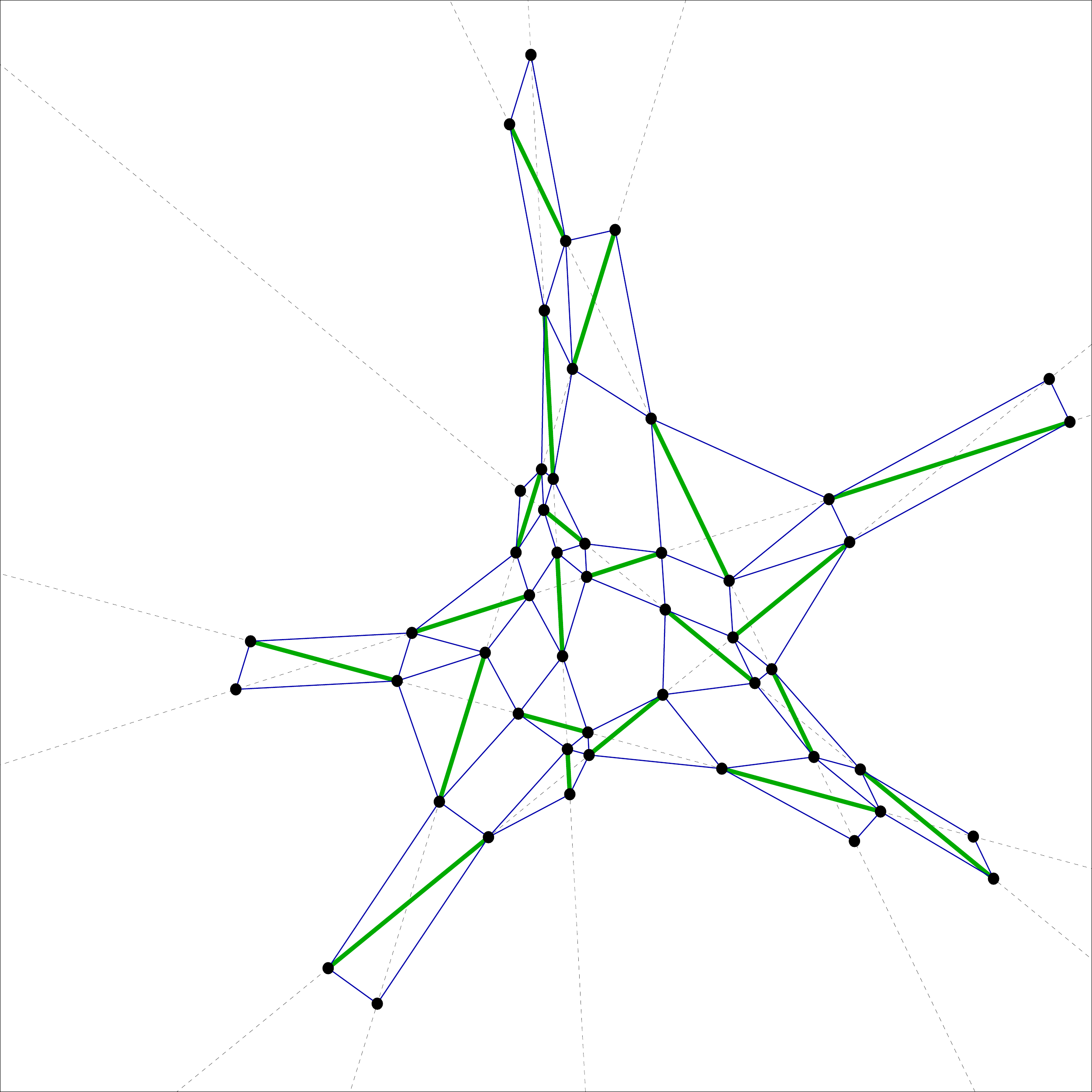}
\end{subfigure}
\hfill
\begin{subfigure}[b]{0.45\textwidth}
\caption{}\label{fig-S5}
\centering
\includegraphics[width=\textwidth]{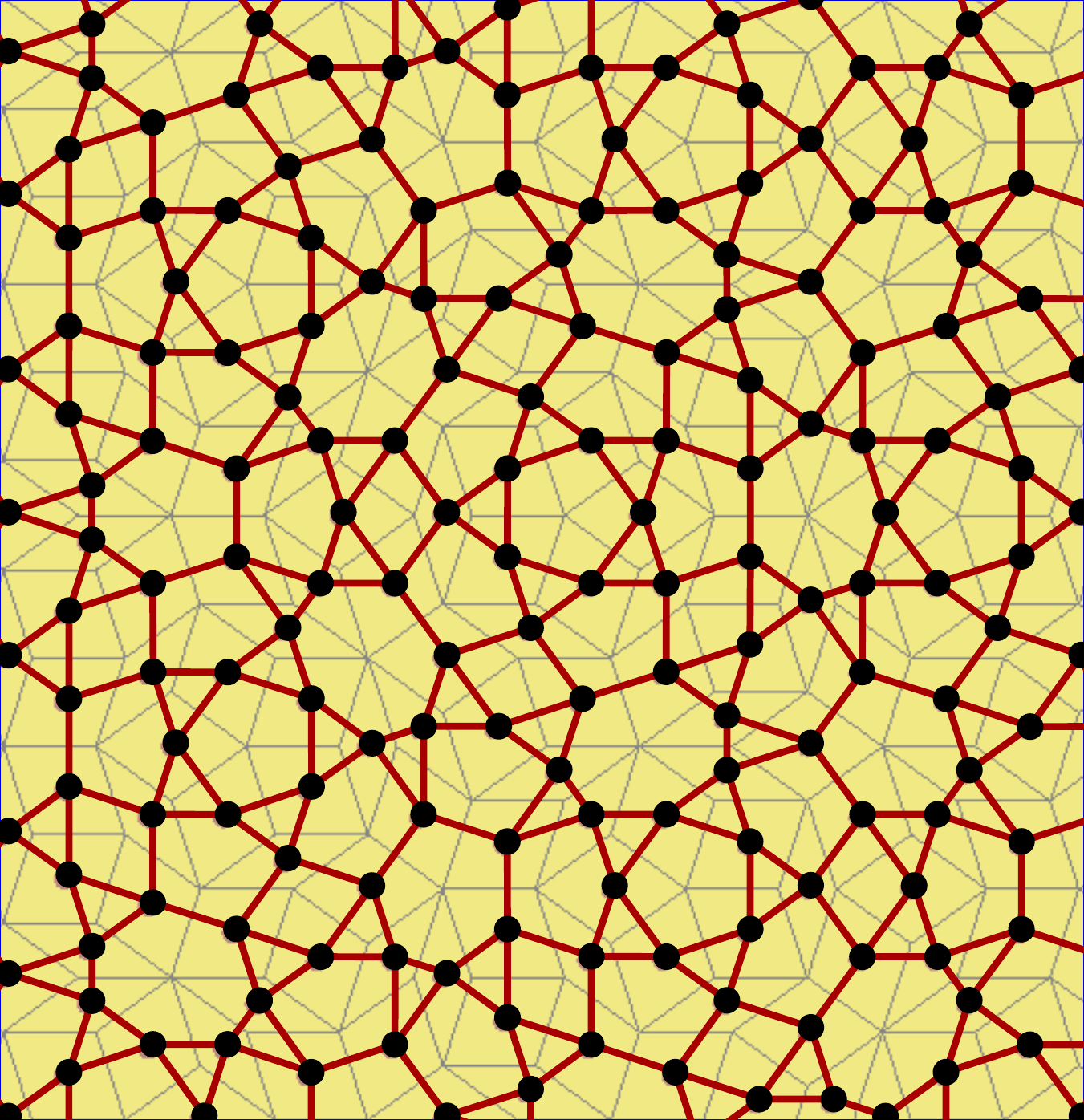}
\end{subfigure}
\caption{(a) An example of a amorphous system with exact ground state constructed via the prescription put forward in the main text. The dashed lines are placed randomly on the plane to create an amorphous system with coordination number 4. {(b) An example of a four coordinated quasicrystal (foreground) generated from Penrose rhomb tiling (background). The method explained in the main text now can de directly used on this system to produce a model with an exact dimer eigen state.}}
\label{fig:child_23 (RIght)}
\end{figure*}

In the main text we have mentioned that a four coordinated quasicrystal can be gnerated from an 
existing quasicrystal made up of quadrangles, e.g. the Penrose rhomb tiling~\cite{Penrose1974,Grunbaum1977} -- one places a vertex in the middle of each quadrangular tile, and then connects all pairs of vertices if
their corresponding tiles share an edge. Fig.~\ref{fig-S5} shows the example of such a quasicrystal with coordination number 4 generated from the Penrose rhomb tiling. The method explained in the main text now can de directly used on the system.

\begin{thebibliography}{48}%
\makeatletter
\providecommand \@ifxundefined [1]{%
 \@ifx{#1\undefined}
}%
\providecommand \@ifnum [1]{%
 \ifnum #1\expandafter \@firstoftwo
 \else \expandafter \@secondoftwo
 \fi
}%
\providecommand \@ifx [1]{%
 \ifx #1\expandafter \@firstoftwo
 \else \expandafter \@secondoftwo
 \fi
}%
\providecommand \natexlab [1]{#1}%
\providecommand \enquote  [1]{``#1''}%
\providecommand \bibnamefont  [1]{#1}%
\providecommand \bibfnamefont [1]{#1}%
\providecommand \citenamefont [1]{#1}%
\providecommand \href@noop [0]{\@secondoftwo}%
\providecommand \href [0]{\begingroup \@sanitize@url \@href}%
\providecommand \@href[1]{\@@startlink{#1}\@@href}%
\providecommand \@@href[1]{\endgroup#1\@@endlink}%
\providecommand \@sanitize@url [0]{\catcode `\\12\catcode `\$12\catcode
  `\&12\catcode `\#12\catcode `\^12\catcode `\_12\catcode `\%12\relax}%
\providecommand \@@startlink[1]{}%
\providecommand \@@endlink[0]{}%
\providecommand \url  [0]{\begingroup\@sanitize@url \@url }%
\providecommand \@url [1]{\endgroup\@href {#1}{\urlprefix }}%
\providecommand \urlprefix  [0]{URL }%
\providecommand \Eprint [0]{\href }%
\providecommand \doibase [0]{http://dx.doi.org/}%
\providecommand \selectlanguage [0]{\@gobble}%
\providecommand \bibinfo  [0]{\@secondoftwo}%
\providecommand \bibfield  [0]{\@secondoftwo}%
\providecommand \translation [1]{[#1]}%
\providecommand \BibitemOpen [0]{}%
\providecommand \bibitemStop [0]{}%
\providecommand \bibitemNoStop [0]{.\EOS\space}%
\providecommand \EOS [0]{\spacefactor3000\relax}%
\providecommand \BibitemShut  [1]{\csname bibitem#1\endcsname}%
\let\auto@bib@innerbib\@empty
\bibitem [{\citenamefont {Lacroix}\ \emph {et~al.}(2011)\citenamefont
  {Lacroix}, \citenamefont {Mendels},\ and\ \citenamefont
  {Mila}}]{frustrationbook}%
  \BibitemOpen
  \bibfield  {author} {\bibinfo {author} {\bibfnamefont {C.}~\bibnamefont
  {Lacroix}}, \bibinfo {author} {\bibfnamefont {P.}~\bibnamefont {Mendels}}, \
  and\ \bibinfo {author} {\bibfnamefont {F.}~\bibnamefont {Mila}},\ }\href
  {https://books.google.de/books?id=utSV09ZuhOkC} {\emph {\bibinfo {title}
  {Introduction to Frustrated Magnetism: Materials, Experiments, Theory}}},\
  Springer Series in Solid-State Sciences\ (\bibinfo  {publisher} {Springer
  Berlin Heidelberg},\ \bibinfo {year} {2011})\BibitemShut {NoStop}%
\bibitem [{\citenamefont {Diep}(2012)}]{diepbook}%
  \BibitemOpen
  \bibfield  {author} {\bibinfo {author} {\bibfnamefont {H.~T.}\ \bibnamefont
  {Diep}},\ }\href {\doibase 10.1142/8676} {\emph {\bibinfo {title} {Frustrated
  Spin Systems}}}\ (\bibinfo  {publisher} {{WORLD} {SCIENTIFIC}},\ \bibinfo
  {year} {2012})\BibitemShut {NoStop}%
\bibitem [{\citenamefont {Wen}\ \emph {et~al.}(1989)\citenamefont {Wen},
  \citenamefont {Wilczek},\ and\ \citenamefont {Zee}}]{Wen1989}%
  \BibitemOpen
  \bibfield  {author} {\bibinfo {author} {\bibfnamefont {X.~G.}\ \bibnamefont
  {Wen}}, \bibinfo {author} {\bibfnamefont {F.}~\bibnamefont {Wilczek}}, \ and\
  \bibinfo {author} {\bibfnamefont {A.}~\bibnamefont {Zee}},\ }\href {\doibase
  10.1103/PhysRevB.39.11413} {\bibfield  {journal} {\bibinfo  {journal} {Phys.
  Rev. B}\ }\textbf {\bibinfo {volume} {39}},\ \bibinfo {pages} {11413}
  (\bibinfo {year} {1989})}\BibitemShut {NoStop}%
\bibitem [{\citenamefont {Wen}(2002)}]{Wen2002}%
  \BibitemOpen
  \bibfield  {author} {\bibinfo {author} {\bibfnamefont {X.-G.}\ \bibnamefont
  {Wen}},\ }\href {\doibase 10.1103/PhysRevB.65.165113} {\bibfield  {journal}
  {\bibinfo  {journal} {Phys. Rev. B}\ }\textbf {\bibinfo {volume} {65}},\
  \bibinfo {pages} {165113} (\bibinfo {year} {2002})}\BibitemShut {NoStop}%
\bibitem [{\citenamefont {Messio}\ \emph {et~al.}(2011)\citenamefont {Messio},
  \citenamefont {Lhuillier},\ and\ \citenamefont {Misguich}}]{Messio2011}%
  \BibitemOpen
  \bibfield  {author} {\bibinfo {author} {\bibfnamefont {L.}~\bibnamefont
  {Messio}}, \bibinfo {author} {\bibfnamefont {C.}~\bibnamefont {Lhuillier}}, \
  and\ \bibinfo {author} {\bibfnamefont {G.}~\bibnamefont {Misguich}},\ }\href
  {\doibase 10.1103/PhysRevB.83.184401} {\bibfield  {journal} {\bibinfo
  {journal} {Phys. Rev. B}\ }\textbf {\bibinfo {volume} {83}},\ \bibinfo
  {pages} {184401} (\bibinfo {year} {2011})}\BibitemShut {NoStop}%
\bibitem [{\citenamefont {Shechtman}\ \emph {et~al.}(1984)\citenamefont
  {Shechtman}, \citenamefont {Blech}, \citenamefont {Gratias},\ and\
  \citenamefont {Cahn}}]{Shechtman1984}%
  \BibitemOpen
  \bibfield  {author} {\bibinfo {author} {\bibfnamefont {D.}~\bibnamefont
  {Shechtman}}, \bibinfo {author} {\bibfnamefont {I.}~\bibnamefont {Blech}},
  \bibinfo {author} {\bibfnamefont {D.}~\bibnamefont {Gratias}}, \ and\
  \bibinfo {author} {\bibfnamefont {J.~W.}\ \bibnamefont {Cahn}},\ }\href
  {\doibase 10.1103/PhysRevLett.53.1951} {\bibfield  {journal} {\bibinfo
  {journal} {Phys. Rev. Lett.}\ }\textbf {\bibinfo {volume} {53}},\ \bibinfo
  {pages} {1951} (\bibinfo {year} {1984})}\BibitemShut {NoStop}%
\bibitem [{\citenamefont {Watanuki}\ \emph {et~al.}(2012)\citenamefont
  {Watanuki}, \citenamefont {Kashimoto}, \citenamefont {Kawana}, \citenamefont
  {Yamazaki}, \citenamefont {Machida}, \citenamefont {Tanaka},\ and\
  \citenamefont {Sato}}]{Watanuki2012}%
  \BibitemOpen
  \bibfield  {author} {\bibinfo {author} {\bibfnamefont {T.}~\bibnamefont
  {Watanuki}}, \bibinfo {author} {\bibfnamefont {S.}~\bibnamefont {Kashimoto}},
  \bibinfo {author} {\bibfnamefont {D.}~\bibnamefont {Kawana}}, \bibinfo
  {author} {\bibfnamefont {T.}~\bibnamefont {Yamazaki}}, \bibinfo {author}
  {\bibfnamefont {A.}~\bibnamefont {Machida}}, \bibinfo {author} {\bibfnamefont
  {Y.}~\bibnamefont {Tanaka}}, \ and\ \bibinfo {author} {\bibfnamefont {T.~J.}\
  \bibnamefont {Sato}},\ }\href {\doibase 10.1103/PhysRevB.86.094201}
  {\bibfield  {journal} {\bibinfo  {journal} {Phys. Rev. B}\ }\textbf {\bibinfo
  {volume} {86}},\ \bibinfo {pages} {094201} (\bibinfo {year}
  {2012})}\BibitemShut {NoStop}%
\bibitem [{\citenamefont {Ishikawa}\ \emph {et~al.}(2016)\citenamefont
  {Ishikawa}, \citenamefont {Hiroto}, \citenamefont {Tokiwa}, \citenamefont
  {Fujii},\ and\ \citenamefont {Tamura}}]{Ishikawa2016}%
  \BibitemOpen
  \bibfield  {author} {\bibinfo {author} {\bibfnamefont {A.}~\bibnamefont
  {Ishikawa}}, \bibinfo {author} {\bibfnamefont {T.}~\bibnamefont {Hiroto}},
  \bibinfo {author} {\bibfnamefont {K.}~\bibnamefont {Tokiwa}}, \bibinfo
  {author} {\bibfnamefont {T.}~\bibnamefont {Fujii}}, \ and\ \bibinfo {author}
  {\bibfnamefont {R.}~\bibnamefont {Tamura}},\ }\href {\doibase
  10.1103/PhysRevB.93.024416} {\bibfield  {journal} {\bibinfo  {journal} {Phys.
  Rev. B}\ }\textbf {\bibinfo {volume} {93}},\ \bibinfo {pages} {024416}
  (\bibinfo {year} {2016})}\BibitemShut {NoStop}%
\bibitem [{\citenamefont {Sato}\ \emph {et~al.}(2019)\citenamefont {Sato},
  \citenamefont {Ishikawa}, \citenamefont {Sakurai}, \citenamefont {Hattori},
  \citenamefont {Avdeev},\ and\ \citenamefont {Tamura}}]{Sato2019}%
  \BibitemOpen
  \bibfield  {author} {\bibinfo {author} {\bibfnamefont {T.~J.}\ \bibnamefont
  {Sato}}, \bibinfo {author} {\bibfnamefont {A.}~\bibnamefont {Ishikawa}},
  \bibinfo {author} {\bibfnamefont {A.}~\bibnamefont {Sakurai}}, \bibinfo
  {author} {\bibfnamefont {M.}~\bibnamefont {Hattori}}, \bibinfo {author}
  {\bibfnamefont {M.}~\bibnamefont {Avdeev}}, \ and\ \bibinfo {author}
  {\bibfnamefont {R.}~\bibnamefont {Tamura}},\ }\href {\doibase
  10.1103/PhysRevB.100.054417} {\bibfield  {journal} {\bibinfo  {journal}
  {Phys. Rev. B}\ }\textbf {\bibinfo {volume} {100}},\ \bibinfo {pages}
  {054417} (\bibinfo {year} {2019})}\BibitemShut {NoStop}%
\bibitem [{\citenamefont {Miyazaki}\ \emph {et~al.}(2020)\citenamefont
  {Miyazaki}, \citenamefont {Sugimoto}, \citenamefont {Morita},\ and\
  \citenamefont {Tohyama}}]{Miyazaki2020}%
  \BibitemOpen
  \bibfield  {author} {\bibinfo {author} {\bibfnamefont {H.}~\bibnamefont
  {Miyazaki}}, \bibinfo {author} {\bibfnamefont {T.}~\bibnamefont {Sugimoto}},
  \bibinfo {author} {\bibfnamefont {K.}~\bibnamefont {Morita}}, \ and\ \bibinfo
  {author} {\bibfnamefont {T.}~\bibnamefont {Tohyama}},\ }\href {\doibase
  10.1103/PhysRevMaterials.4.024417} {\bibfield  {journal} {\bibinfo  {journal}
  {Phys. Rev. Mater.}\ }\textbf {\bibinfo {volume} {4}},\ \bibinfo {pages}
  {024417} (\bibinfo {year} {2020})}\BibitemShut {NoStop}%
\bibitem [{\citenamefont {Janot}(1992)}]{Janot1992}%
  \BibitemOpen
  \bibfield  {author} {\bibinfo {author} {\bibfnamefont {C.}~\bibnamefont
  {Janot}},\ }\href@noop {} {\emph {\bibinfo {title} {Quasicrystals - A
  Primer}}}\ (\bibinfo  {publisher} {Clarendon Press},\ \bibinfo {address}
  {Oxford},\ \bibinfo {year} {1992})\BibitemShut {NoStop}%
\bibitem [{\citenamefont {Fukamichi}\ \emph {et~al.}(1987)\citenamefont
  {Fukamichi}, \citenamefont {Goto}, \citenamefont {Masumoto}, \citenamefont
  {Sakakibara}, \citenamefont {Oguchi},\ and\ \citenamefont
  {Todo}}]{Fukamichi1987}%
  \BibitemOpen
  \bibfield  {author} {\bibinfo {author} {\bibfnamefont {K.}~\bibnamefont
  {Fukamichi}}, \bibinfo {author} {\bibfnamefont {T.}~\bibnamefont {Goto}},
  \bibinfo {author} {\bibfnamefont {T.}~\bibnamefont {Masumoto}}, \bibinfo
  {author} {\bibfnamefont {T.}~\bibnamefont {Sakakibara}}, \bibinfo {author}
  {\bibfnamefont {M.}~\bibnamefont {Oguchi}}, \ and\ \bibinfo {author}
  {\bibfnamefont {S.}~\bibnamefont {Todo}},\ }\href {\doibase
  10.1088/0305-4608/17/3/018} {\bibfield  {journal} {\bibinfo  {journal}
  {Journal of Physics F: Metal Physics}\ }\textbf {\bibinfo {volume} {17}},\
  \bibinfo {pages} {743} (\bibinfo {year} {1987})}\BibitemShut {NoStop}%
\bibitem [{\citenamefont {Berger}\ and\ \citenamefont
  {Prejean}(1990)}]{Berger1990}%
  \BibitemOpen
  \bibfield  {author} {\bibinfo {author} {\bibfnamefont {C.}~\bibnamefont
  {Berger}}\ and\ \bibinfo {author} {\bibfnamefont {J.~J.}\ \bibnamefont
  {Prejean}},\ }\href {\doibase 10.1103/physrevlett.64.1769} {\bibfield
  {journal} {\bibinfo  {journal} {Physical Review Letters}\ }\textbf {\bibinfo
  {volume} {64}},\ \bibinfo {pages} {1769} (\bibinfo {year}
  {1990})}\BibitemShut {NoStop}%
\bibitem [{\citenamefont {Islam}\ \emph {et~al.}(1998)\citenamefont {Islam},
  \citenamefont {Fisher}, \citenamefont {Zarestky}, \citenamefont {Canfield},
  \citenamefont {Stassis},\ and\ \citenamefont {Goldman}}]{Islam1998}%
  \BibitemOpen
  \bibfield  {author} {\bibinfo {author} {\bibfnamefont {Z.}~\bibnamefont
  {Islam}}, \bibinfo {author} {\bibfnamefont {I.~R.}\ \bibnamefont {Fisher}},
  \bibinfo {author} {\bibfnamefont {J.}~\bibnamefont {Zarestky}}, \bibinfo
  {author} {\bibfnamefont {P.~C.}\ \bibnamefont {Canfield}}, \bibinfo {author}
  {\bibfnamefont {C.}~\bibnamefont {Stassis}}, \ and\ \bibinfo {author}
  {\bibfnamefont {A.~I.}\ \bibnamefont {Goldman}},\ }\href {\doibase
  10.1103/PhysRevB.57.R11047} {\bibfield  {journal} {\bibinfo  {journal} {Phys.
  Rev. B}\ }\textbf {\bibinfo {volume} {57}},\ \bibinfo {pages} {R11047}
  (\bibinfo {year} {1998})}\BibitemShut {NoStop}%
\bibitem [{\citenamefont {Shastry}\ and\ \citenamefont
  {Sutherland}(1981)}]{Shastry1981}%
  \BibitemOpen
  \bibfield  {author} {\bibinfo {author} {\bibfnamefont {B.~S.}\ \bibnamefont
  {Shastry}}\ and\ \bibinfo {author} {\bibfnamefont {B.}~\bibnamefont
  {Sutherland}},\ }\href {\doibase 10.1016/0378-4363(81)90838-x} {\bibfield
  {journal} {\bibinfo  {journal} {Physica B+C}\ }\textbf {\bibinfo {volume}
  {108}},\ \bibinfo {pages} {1069} (\bibinfo {year} {1981})}\BibitemShut
  {NoStop}%
\bibitem [{\citenamefont {Ghosh}\ \emph {et~al.}(2022)\citenamefont {Ghosh},
  \citenamefont {M\"uller},\ and\ \citenamefont {Thomale}}]{Ghosh2022}%
  \BibitemOpen
  \bibfield  {author} {\bibinfo {author} {\bibfnamefont {P.}~\bibnamefont
  {Ghosh}}, \bibinfo {author} {\bibfnamefont {T.}~\bibnamefont {M\"uller}}, \
  and\ \bibinfo {author} {\bibfnamefont {R.}~\bibnamefont {Thomale}},\ }\href
  {\doibase 10.1103/PhysRevB.105.L180412} {\bibfield  {journal} {\bibinfo
  {journal} {Phys. Rev. B}\ }\textbf {\bibinfo {volume} {105}},\ \bibinfo
  {pages} {L180412} (\bibinfo {year} {2022})}\BibitemShut {NoStop}%
\bibitem [{\citenamefont {Korepin}(1987)}]{Korepin1987}%
  \BibitemOpen
  \bibfield  {author} {\bibinfo {author} {\bibfnamefont {V.~E.}\ \bibnamefont
  {Korepin}},\ }\href {\doibase 10.1007/bf01209021} {\bibfield  {journal}
  {\bibinfo  {journal} {Communications in Mathematical Physics}\ }\textbf
  {\bibinfo {volume} {110}},\ \bibinfo {pages} {157} (\bibinfo {year}
  {1987})}\BibitemShut {NoStop}%
\bibitem [{\citenamefont {Jeon}\ \emph {et~al.}(2022)\citenamefont {Jeon},
  \citenamefont {Kim},\ and\ \citenamefont {Lee}}]{Jeon2022}%
  \BibitemOpen
  \bibfield  {author} {\bibinfo {author} {\bibfnamefont {J.}~\bibnamefont
  {Jeon}}, \bibinfo {author} {\bibfnamefont {S.~K.}\ \bibnamefont {Kim}}, \
  and\ \bibinfo {author} {\bibfnamefont {S.~B.}~\bibnamefont {Lee}},\ }\href
  {\doibase 10.1103/PhysRevB.106.134431} {\bibfield  {journal} {\bibinfo
  {journal} {Phys. Rev. B}\ }\textbf {\bibinfo {volume} {106}},\ \bibinfo
  {pages} {134431} (\bibinfo {year} {2022})}\BibitemShut {NoStop}%
\bibitem [{\citenamefont {Hauser}\ \emph {et~al.}(1986)\citenamefont {Hauser},
  \citenamefont {Chen},\ and\ \citenamefont {Waszczak}}]{Hauser1986}%
  \BibitemOpen
  \bibfield  {author} {\bibinfo {author} {\bibfnamefont {J.~J.}\ \bibnamefont
  {Hauser}}, \bibinfo {author} {\bibfnamefont {H.~S.}\ \bibnamefont {Chen}}, \
  and\ \bibinfo {author} {\bibfnamefont {J.~V.}\ \bibnamefont {Waszczak}},\
  }\href {\doibase 10.1103/PhysRevB.33.3577} {\bibfield  {journal} {\bibinfo
  {journal} {Phys. Rev. B}\ }\textbf {\bibinfo {volume} {33}},\ \bibinfo
  {pages} {3577} (\bibinfo {year} {1986})}\BibitemShut {NoStop}%
\bibitem [{\citenamefont {Oxborrow}\ and\ \citenamefont
  {Henley}(1993)}]{Oxborrow1993}%
  \BibitemOpen
  \bibfield  {author} {\bibinfo {author} {\bibfnamefont {M.}~\bibnamefont
  {Oxborrow}}\ and\ \bibinfo {author} {\bibfnamefont {C.~L.}\ \bibnamefont
  {Henley}},\ }\href {\doibase 10.1103/PhysRevB.48.6966} {\bibfield  {journal}
  {\bibinfo  {journal} {Phys. Rev. B}\ }\textbf {\bibinfo {volume} {48}},\
  \bibinfo {pages} {6966} (\bibinfo {year} {1993})}\BibitemShut {NoStop}%
\bibitem [{\citenamefont {Wessel}\ \emph {et~al.}(2003)\citenamefont {Wessel},
  \citenamefont {Jagannathan},\ and\ \citenamefont {Haas}}]{Wessel2003}%
  \BibitemOpen
  \bibfield  {author} {\bibinfo {author} {\bibfnamefont {S.}~\bibnamefont
  {Wessel}}, \bibinfo {author} {\bibfnamefont {A.}~\bibnamefont {Jagannathan}},
  \ and\ \bibinfo {author} {\bibfnamefont {S.}~\bibnamefont {Haas}},\ }\href
  {\doibase 10.1103/PhysRevLett.90.177205} {\bibfield  {journal} {\bibinfo
  {journal} {Phys. Rev. Lett.}\ }\textbf {\bibinfo {volume} {90}},\ \bibinfo
  {pages} {177205} (\bibinfo {year} {2003})}\BibitemShut {NoStop}%
\bibitem [{\citenamefont {Wessel}\ and\ \citenamefont
  {Milat}(2005)}]{Wessel2005}%
  \BibitemOpen
  \bibfield  {author} {\bibinfo {author} {\bibfnamefont {S.}~\bibnamefont
  {Wessel}}\ and\ \bibinfo {author} {\bibfnamefont {I.}~\bibnamefont {Milat}},\
  }\href {\doibase 10.1103/PhysRevB.71.104427} {\bibfield  {journal} {\bibinfo
  {journal} {Phys. Rev. B}\ }\textbf {\bibinfo {volume} {71}},\ \bibinfo
  {pages} {104427} (\bibinfo {year} {2005})}\BibitemShut {NoStop}%
\bibitem [{\citenamefont {Jagannathan}(2005)}]{Jagannathan2005}%
  \BibitemOpen
  \bibfield  {author} {\bibinfo {author} {\bibfnamefont {A.}~\bibnamefont
  {Jagannathan}},\ }\href {\doibase 10.1103/PhysRevB.71.115101} {\bibfield
  {journal} {\bibinfo  {journal} {Phys. Rev. B}\ }\textbf {\bibinfo {volume}
  {71}},\ \bibinfo {pages} {115101} (\bibinfo {year} {2005})}\BibitemShut
  {NoStop}%
\bibitem [{\citenamefont {Jagannathan}\ \emph {et~al.}(2007)\citenamefont
  {Jagannathan}, \citenamefont {Szallas}, \citenamefont {Wessel},\ and\
  \citenamefont {Duneau}}]{Jagannathan2007}%
  \BibitemOpen
  \bibfield  {author} {\bibinfo {author} {\bibfnamefont {A.}~\bibnamefont
  {Jagannathan}}, \bibinfo {author} {\bibfnamefont {A.}~\bibnamefont
  {Szallas}}, \bibinfo {author} {\bibfnamefont {S.}~\bibnamefont {Wessel}}, \
  and\ \bibinfo {author} {\bibfnamefont {M.}~\bibnamefont {Duneau}},\ }\href
  {\doibase 10.1103/PhysRevB.75.212407} {\bibfield  {journal} {\bibinfo
  {journal} {Phys. Rev. B}\ }\textbf {\bibinfo {volume} {75}},\ \bibinfo
  {pages} {212407} (\bibinfo {year} {2007})}\BibitemShut {NoStop}%
\bibitem [{\citenamefont {Dotera}\ \emph {et~al.}(2017)\citenamefont {Dotera},
  \citenamefont {Bekku},\ and\ \citenamefont {Ziherl}}]{Dotera2017}%
  \BibitemOpen
  \bibfield  {author} {\bibinfo {author} {\bibfnamefont {T.}~\bibnamefont
  {Dotera}}, \bibinfo {author} {\bibfnamefont {S.}~\bibnamefont {Bekku}}, \
  and\ \bibinfo {author} {\bibfnamefont {P.}~\bibnamefont {Ziherl}},\ }\href
  {\doibase 10.1038/nmat4963} {\bibfield  {journal} {\bibinfo  {journal}
  {Nature Materials}\ }\textbf {\bibinfo {volume} {16}},\ \bibinfo {pages}
  {987} (\bibinfo {year} {2017})}\BibitemShut {NoStop}%
\bibitem [{\citenamefont {Penrose}(1974)}]{Penrose1974}%
  \BibitemOpen
  \bibfield  {author} {\bibinfo {author} {\bibfnamefont {R.}~\bibnamefont
  {Penrose}},\ }\href@noop {} {\bibfield  {journal} {\bibinfo  {journal}
  {Bulletin of the Institute of Mathematics and Its Applications}\ ,\ \bibinfo
  {pages} {266}} (\bibinfo {year} {1974})}\BibitemShut {NoStop}%
\bibitem [{\citenamefont {Grunbaum}\ and\ \citenamefont
  {Shephard}(1977)}]{Grunbaum1977}%
  \BibitemOpen
  \bibfield  {author} {\bibinfo {author} {\bibfnamefont {B.}~\bibnamefont
  {Grunbaum}}\ and\ \bibinfo {author} {\bibfnamefont {G.~C.}\ \bibnamefont
  {Shephard}},\ }\href {http://www.jstor.org/stable/2689529} {\bibfield
  {journal} {\bibinfo  {journal} {Mathematics Magazine}\ }\textbf {\bibinfo
  {volume} {50}},\ \bibinfo {pages} {227} (\bibinfo {year} {1977})}\BibitemShut
  {NoStop}%
\bibitem [{\citenamefont {Majumdar}\ and\ \citenamefont
  {Ghosh}(1969)}]{Majumdar1969}%
  \BibitemOpen
  \bibfield  {author} {\bibinfo {author} {\bibfnamefont {C.~K.}\ \bibnamefont
  {Majumdar}}\ and\ \bibinfo {author} {\bibfnamefont {D.~K.}\ \bibnamefont
  {Ghosh}},\ }\href {\doibase 10.1063/1.1664978} {\bibfield  {journal}
  {\bibinfo  {journal} {J. Math. Phys.}\ }\textbf {\bibinfo {volume} {10}},\
  \bibinfo {pages} {1388} (\bibinfo {year} {1969})}\BibitemShut {NoStop}%
\bibitem [{\citenamefont {Affleck}\ \emph {et~al.}(1987)\citenamefont
  {Affleck}, \citenamefont {Kennedy}, \citenamefont {Lieb},\ and\ \citenamefont
  {Tasaki}}]{AKLT}%
  \BibitemOpen
  \bibfield  {author} {\bibinfo {author} {\bibfnamefont {I.}~\bibnamefont
  {Affleck}}, \bibinfo {author} {\bibfnamefont {T.}~\bibnamefont {Kennedy}},
  \bibinfo {author} {\bibfnamefont {E.~H.}\ \bibnamefont {Lieb}}, \ and\
  \bibinfo {author} {\bibfnamefont {H.}~\bibnamefont {Tasaki}},\ }\href
  {\doibase 10.1103/PhysRevLett.59.799} {\bibfield  {journal} {\bibinfo
  {journal} {Phys. Rev. Lett.}\ }\textbf {\bibinfo {volume} {59}},\ \bibinfo
  {pages} {799} (\bibinfo {year} {1987})}\BibitemShut {NoStop}%
\bibitem [{\citenamefont {Klein}(1982)}]{Klein_1982}%
  \BibitemOpen
  \bibfield  {author} {\bibinfo {author} {\bibfnamefont {D.~J.}\ \bibnamefont
  {Klein}},\ }\href {\doibase 10.1088/0305-4470/15/2/032} {\bibfield  {journal}
  {\bibinfo  {journal} {Journal of Physics A: Mathematical and General}\
  }\textbf {\bibinfo {volume} {15}},\ \bibinfo {pages} {661} (\bibinfo {year}
  {1982})}\BibitemShut {NoStop}%
\bibitem [{Note1()}]{Note1}%
  \BibitemOpen
  \bibinfo {note} {Here, we only covered Heisenberg spin interactions. In
  general, an analysis along these lines can also be achieved for XXZ-type
  interactions and a Zeeman term. For further details, see Refs.~\cite
  {Shastry1981, Ghosh2022}}\BibitemShut {NoStop}%
\bibitem [{\citenamefont {Corboz}\ and\ \citenamefont
  {Mila}(2013)}]{Corboz2013}%
  \BibitemOpen
  \bibfield  {author} {\bibinfo {author} {\bibfnamefont {P.}~\bibnamefont
  {Corboz}}\ and\ \bibinfo {author} {\bibfnamefont {F.}~\bibnamefont {Mila}},\
  }\href {\doibase 10.1103/PhysRevB.87.115144} {\bibfield  {journal} {\bibinfo
  {journal} {Phys. Rev. B}\ }\textbf {\bibinfo {volume} {87}},\ \bibinfo
  {pages} {115144} (\bibinfo {year} {2013})}\BibitemShut {NoStop}%
\bibitem [{\citenamefont {Lee}\ \emph {et~al.}(2019)\citenamefont {Lee},
  \citenamefont {You}, \citenamefont {Sachdev},\ and\ \citenamefont
  {Vishwanath}}]{Lee2019}%
  \BibitemOpen
  \bibfield  {author} {\bibinfo {author} {\bibfnamefont {J.~Y.}\ \bibnamefont
  {Lee}}, \bibinfo {author} {\bibfnamefont {Y.-Z.}\ \bibnamefont {You}},
  \bibinfo {author} {\bibfnamefont {S.}~\bibnamefont {Sachdev}}, \ and\
  \bibinfo {author} {\bibfnamefont {A.}~\bibnamefont {Vishwanath}},\ }\href
  {\doibase 10.1103/PhysRevX.9.041037} {\bibfield  {journal} {\bibinfo
  {journal} {Phys. Rev. X}\ }\textbf {\bibinfo {volume} {9}},\ \bibinfo {pages}
  {041037} (\bibinfo {year} {2019})}\BibitemShut {NoStop}%
\bibitem [{\citenamefont {Farnell}\ \emph {et~al.}(2011)\citenamefont
  {Farnell}, \citenamefont {Darradi}, \citenamefont {Schmidt},\ and\
  \citenamefont {Richter}}]{Farnell2011}%
  \BibitemOpen
  \bibfield  {author} {\bibinfo {author} {\bibfnamefont {D.~J.~J.}\
  \bibnamefont {Farnell}}, \bibinfo {author} {\bibfnamefont {R.}~\bibnamefont
  {Darradi}}, \bibinfo {author} {\bibfnamefont {R.}~\bibnamefont {Schmidt}}, \
  and\ \bibinfo {author} {\bibfnamefont {J.}~\bibnamefont {Richter}},\ }\href
  {\doibase 10.1103/PhysRevB.84.104406} {\bibfield  {journal} {\bibinfo
  {journal} {Phys. Rev. B}\ }\textbf {\bibinfo {volume} {84}},\ \bibinfo
  {pages} {104406} (\bibinfo {year} {2011})}\BibitemShut {NoStop}%
\bibitem [{\citenamefont {White}(1992)}]{White1992}%
  \BibitemOpen
  \bibfield  {author} {\bibinfo {author} {\bibfnamefont {S.~R.}\ \bibnamefont
  {White}},\ }\href {\doibase 10.1103/PhysRevLett.69.2863} {\bibfield
  {journal} {\bibinfo  {journal} {Phys. Rev. Lett.}\ }\textbf {\bibinfo
  {volume} {69}},\ \bibinfo {pages} {2863} (\bibinfo {year}
  {1992})}\BibitemShut {NoStop}%
\bibitem [{\citenamefont {Fishman}\ \emph {et~al.}(2020)\citenamefont
  {Fishman}, \citenamefont {White},\ and\ \citenamefont
  {Stoudenmire}}]{Fishman2020}%
  \BibitemOpen
  \bibfield  {author} {\bibinfo {author} {\bibfnamefont {M.}~\bibnamefont
  {Fishman}}, \bibinfo {author} {\bibfnamefont {S.~R.}\ \bibnamefont {White}},
  \ and\ \bibinfo {author} {\bibfnamefont {E.~M.}\ \bibnamefont
  {Stoudenmire}},\ }\href@noop {} {\enquote {\bibinfo {title} {The itensor
  software library for tensor network calculations},}\ } (\bibinfo {year}
  {2020}),\ \Eprint {http://arxiv.org/abs/2007.14822} {arXiv:2007.14822
  [cs.MS]} \BibitemShut {NoStop}%
\bibitem [{\citenamefont {Senthil}\ \emph {et~al.}(2004)\citenamefont
  {Senthil}, \citenamefont {Vishwanath}, \citenamefont {Balents}, \citenamefont
  {Sachdev},\ and\ \citenamefont {Fisher}}]{deconfined}%
  \BibitemOpen
  \bibfield  {author} {\bibinfo {author} {\bibfnamefont {T.}~\bibnamefont
  {Senthil}}, \bibinfo {author} {\bibfnamefont {A.}~\bibnamefont {Vishwanath}},
  \bibinfo {author} {\bibfnamefont {L.}~\bibnamefont {Balents}}, \bibinfo
  {author} {\bibfnamefont {S.}~\bibnamefont {Sachdev}}, \ and\ \bibinfo
  {author} {\bibfnamefont {M.~P.~A.}\ \bibnamefont {Fisher}},\ }\href {\doibase
  10.1126/science.1091806} {\bibfield  {journal} {\bibinfo  {journal}
  {Science}\ }\textbf {\bibinfo {volume} {303}},\ \bibinfo {pages} {1490}
  (\bibinfo {year} {2004})}\BibitemShut {NoStop}%
\bibitem [{\citenamefont {Shi}\ \emph {et~al.}(2022)\citenamefont {Shi},
  \citenamefont {Dissanayake}, \citenamefont {Corboz}, \citenamefont
  {Steinhardt}, \citenamefont {Graf}, \citenamefont {Silevitch}, \citenamefont
  {Dabkowska}, \citenamefont {Rosenbaum}, \citenamefont {Mila},\ and\
  \citenamefont {Haravifard}}]{Shi2022}%
  \BibitemOpen
  \bibfield  {author} {\bibinfo {author} {\bibfnamefont {Z.}~\bibnamefont
  {Shi}}, \bibinfo {author} {\bibfnamefont {S.}~\bibnamefont {Dissanayake}},
  \bibinfo {author} {\bibfnamefont {P.}~\bibnamefont {Corboz}}, \bibinfo
  {author} {\bibfnamefont {W.}~\bibnamefont {Steinhardt}}, \bibinfo {author}
  {\bibfnamefont {D.}~\bibnamefont {Graf}}, \bibinfo {author} {\bibfnamefont
  {D.~M.}\ \bibnamefont {Silevitch}}, \bibinfo {author} {\bibfnamefont {H.~A.}\
  \bibnamefont {Dabkowska}}, \bibinfo {author} {\bibfnamefont {T.~F.}\
  \bibnamefont {Rosenbaum}}, \bibinfo {author} {\bibfnamefont {F.}~\bibnamefont
  {Mila}}, \ and\ \bibinfo {author} {\bibfnamefont {S.}~\bibnamefont
  {Haravifard}},\ }\href {\doibase 10.1038/s41467-022-30036-w} {\bibfield
  {journal} {\bibinfo  {journal} {Nature Communications}\ }\textbf {\bibinfo
  {volume} {13}} (\bibinfo {year} {2022}),\
  10.1038/s41467-022-30036-w}\BibitemShut {NoStop}%
\bibitem [{\citenamefont {Yang}\ \emph {et~al.}(2022)\citenamefont {Yang},
  \citenamefont {Sandvik},\ and\ \citenamefont {Wang}}]{Yang2022}%
  \BibitemOpen
  \bibfield  {author} {\bibinfo {author} {\bibfnamefont {J.}~\bibnamefont
  {Yang}}, \bibinfo {author} {\bibfnamefont {A.~W.}\ \bibnamefont {Sandvik}}, \
  and\ \bibinfo {author} {\bibfnamefont {L.}~\bibnamefont {Wang}},\ }\href
  {\doibase 10.1103/PhysRevB.105.L060409} {\bibfield  {journal} {\bibinfo
  {journal} {Phys. Rev. B}\ }\textbf {\bibinfo {volume} {105}},\ \bibinfo
  {pages} {L060409} (\bibinfo {year} {2022})}\BibitemShut {NoStop}%
\bibitem [{\citenamefont {Dotera}\ and\ \citenamefont
  {Abe}(1990)}]{Dotera1990}%
  \BibitemOpen
  \bibfield  {author} {\bibinfo {author} {\bibfnamefont {T.}~\bibnamefont
  {Dotera}}\ and\ \bibinfo {author} {\bibfnamefont {R.}~\bibnamefont {Abe}},\
  }\href {\doibase 10.1143/jpsj.59.2064} {\bibfield  {journal} {\bibinfo
  {journal} {Journal of the Physical Society of Japan}\ }\textbf {\bibinfo
  {volume} {59}},\ \bibinfo {pages} {2064} (\bibinfo {year}
  {1990})}\BibitemShut {NoStop}%
\bibitem [{\citenamefont {Momoi}\ and\ \citenamefont
  {Totsuka}(2000)}]{Momoi2000}%
  \BibitemOpen
  \bibfield  {author} {\bibinfo {author} {\bibfnamefont {T.}~\bibnamefont
  {Momoi}}\ and\ \bibinfo {author} {\bibfnamefont {K.}~\bibnamefont
  {Totsuka}},\ }\href {\doibase 10.1103/PhysRevB.61.3231} {\bibfield  {journal}
  {\bibinfo  {journal} {Phys. Rev. B}\ }\textbf {\bibinfo {volume} {61}},\
  \bibinfo {pages} {3231} (\bibinfo {year} {2000})}\BibitemShut {NoStop}%
\bibitem [{\citenamefont {Fukumoto}\ and\ \citenamefont
  {Oguchi}(2000)}]{Fukumoto2000}%
  \BibitemOpen
  \bibfield  {author} {\bibinfo {author} {\bibfnamefont {Y.}~\bibnamefont
  {Fukumoto}}\ and\ \bibinfo {author} {\bibfnamefont {A.}~\bibnamefont
  {Oguchi}},\ }\href {\doibase 10.1143/jpsj.69.1286} {\bibfield  {journal}
  {\bibinfo  {journal} {J. Phys. Soc. Jpn.}\ }\textbf {\bibinfo {volume}
  {69}},\ \bibinfo {pages} {1286} (\bibinfo {year} {2000})}\BibitemShut
  {NoStop}%
\bibitem [{\citenamefont {Miyahara}\ \emph {et~al.}(2003)\citenamefont
  {Miyahara}, \citenamefont {Becca},\ and\ \citenamefont
  {Mila}}]{Miyahara2003}%
  \BibitemOpen
  \bibfield  {author} {\bibinfo {author} {\bibfnamefont {S.}~\bibnamefont
  {Miyahara}}, \bibinfo {author} {\bibfnamefont {F.}~\bibnamefont {Becca}}, \
  and\ \bibinfo {author} {\bibfnamefont {F.}~\bibnamefont {Mila}},\ }\href
  {\doibase 10.1103/PhysRevB.68.024401} {\bibfield  {journal} {\bibinfo
  {journal} {Phys. Rev. B}\ }\textbf {\bibinfo {volume} {68}},\ \bibinfo
  {pages} {024401} (\bibinfo {year} {2003})}\BibitemShut {NoStop}%
\bibitem [{\citenamefont {Dorier}\ \emph {et~al.}(2008)\citenamefont {Dorier},
  \citenamefont {Schmidt},\ and\ \citenamefont {Mila}}]{Dorier2008}%
  \BibitemOpen
  \bibfield  {author} {\bibinfo {author} {\bibfnamefont {J.}~\bibnamefont
  {Dorier}}, \bibinfo {author} {\bibfnamefont {K.~P.}\ \bibnamefont {Schmidt}},
  \ and\ \bibinfo {author} {\bibfnamefont {F.}~\bibnamefont {Mila}},\ }\href
  {\doibase 10.1103/PhysRevLett.101.250402} {\bibfield  {journal} {\bibinfo
  {journal} {Phys. Rev. Lett.}\ }\textbf {\bibinfo {volume} {101}},\ \bibinfo
  {pages} {250402} (\bibinfo {year} {2008})}\BibitemShut {NoStop}%
\bibitem [{\citenamefont {Corboz}\ and\ \citenamefont
  {Mila}(2014)}]{Corboz2014}%
  \BibitemOpen
  \bibfield  {author} {\bibinfo {author} {\bibfnamefont {P.}~\bibnamefont
  {Corboz}}\ and\ \bibinfo {author} {\bibfnamefont {F.}~\bibnamefont {Mila}},\
  }\href {\doibase 10.1103/PhysRevLett.112.147203} {\bibfield  {journal}
  {\bibinfo  {journal} {Phys. Rev. Lett.}\ }\textbf {\bibinfo {volume} {112}},\
  \bibinfo {pages} {147203} (\bibinfo {year} {2014})}\BibitemShut {NoStop}%
  \bibitem [{\citenamefont {Koga}\ and\ \citenamefont
  {Kawakami}(2014)}]{Koga2000}%
  \BibitemOpen
  \bibfield  {author} {\bibinfo {author} {\bibfnamefont {A.}~\bibnamefont
  {Koga}}\ and\ \bibinfo {author} {\bibfnamefont {N.}~\bibnamefont {Kawakami}},\
  }\href {\doibase 10.1103/PhysRevLett.84.4461} {\bibfield  {journal}
  {\bibinfo  {journal} {Phys. Rev. Lett.}\ }\textbf {\bibinfo {volume} {84}},\
  \bibinfo {pages} {4461} (\bibinfo {year} {2000})}\BibitemShut {NoStop}%
\bibitem [{\citenamefont {Ghosh}\ \emph {et~al.}()\citenamefont {Ghosh},
  \citenamefont {Seufert}, \citenamefont {Müller}, \citenamefont {Mila},\ and\
  \citenamefont {Thomale}}]{Ghosh2023}%
  \BibitemOpen
  \bibfield  {author} {\bibinfo {author} {\bibfnamefont {P.}~\bibnamefont
  {Ghosh}}, \bibinfo {author} {\bibfnamefont {J.}~\bibnamefont {Seufert}},
  \bibinfo {author} {\bibfnamefont {T.}~\bibnamefont {Müller}}, \bibinfo
  {author} {\bibfnamefont {F.}~\bibnamefont {Mila}}, \ and\ \bibinfo {author}
  {\bibfnamefont {R.}~\bibnamefont {Thomale}},\ }\href@noop {} {\ }\Eprint
  {http://arxiv.org/abs/2301.08264v1} {2301.08264v1} \BibitemShut {NoStop}%
\bibitem [{\citenamefont {Cassella}\ \emph {et~al.}()\citenamefont {Cassella},
  \citenamefont {D'Ornellas}, \citenamefont {Hodson}, \citenamefont {Natori},\
  and\ \citenamefont {Knolle}}]{Cassella2022}%
  \BibitemOpen
  \bibfield  {author} {\bibinfo {author} {\bibfnamefont {G.}~\bibnamefont
  {Cassella}}, \bibinfo {author} {\bibfnamefont {P.}~\bibnamefont
  {D'Ornellas}}, \bibinfo {author} {\bibfnamefont {T.}~\bibnamefont {Hodson}},
  \bibinfo {author} {\bibfnamefont {W.~M.~H.}\ \bibnamefont {Natori}}, \ and\
  \bibinfo {author} {\bibfnamefont {J.}~\bibnamefont {Knolle}},\ }\href@noop {}
  {\ }\Eprint {http://arxiv.org/abs/2208.08246v2} {2208.08246v2} \BibitemShut
  {NoStop}%
\end{thebibliography}
%

\end{document}